- Perspective –

# Supersolidity of the Confined and the Hydrating Water

Chang Q Sun[1*]

Abstract

Supersolid water, firstly defined in 2013 [*J Phys Chem Letter* **4**: 2565; *ibid* **4**: 3238] and intensively verified since then, refers to those water molecules being polarized by molecular undercoordination (often called confinement such as nanobubbles and droplets) or by salt hydration. This work shows that a combination of the STM/S, XPS, NEXFAS, SFG, DPS, ultrafast UPS, and ultrafast FTIR observations and quantum theory calculations confirmed the bond−electron−phonon correlation in the supersolid phase. The supersolidity is characterized by the shorter and stiffer H−O bond and longer and softer O:H nonbond, O 1s energy entrapment, hydrated electron polarization and the longer lifetime of photoelectrons and phonons. The supersolid phase is less dense, elastoviscous, mechanically and thermally more stable with the hydrophobic and frictionless surface. The O:H−O bond cooperative relaxation disperses outwardly the quasisolid phase boundary to raise the melting point and meanwhile lower the freezing temperature of the quasisolid phase – called supercooling and superheating.

Highlight

- *Molecular undercoordination and ionic hydration effect the same on O:H−O relaxation*
- *XPS O 1s and K−edge absorption energy shifts in proportional to the H−O bond energy*
- *Electron hydration probes the site− and size−resolved bounding energy and the electron lifetime*
- *DPS, SFG, and calculations confirm the site−resolved H−O contraction and thermal stability*

[1] NOVITUS, EEE, Nanyang Technological University, Singapore 639798 (ecqsun@ntu.edu.sg); BEAM, Yangtze Normal University, Chongqing 408100, China (20161042@yznu.cn; ecqsun@qq.com)

Content entry

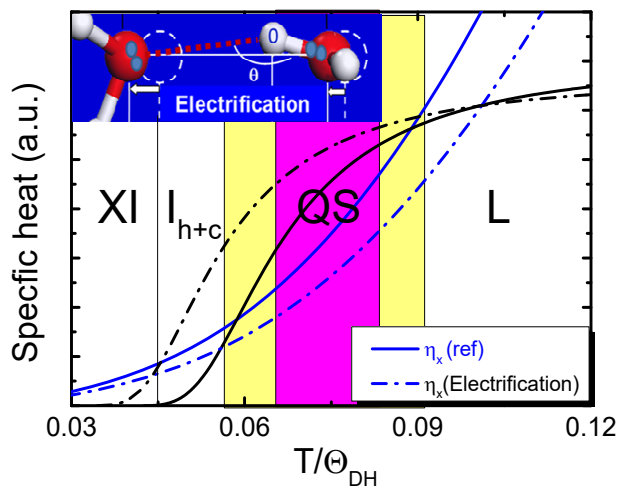

# Contents

1. Wonders of H₂O Molecular Undercoordination and Hydration

As independent degrees of freedom, water molecular undercoordination (also called confinement) and electrostatic polarization make the mysterious water ice even more fascinating [1]. Undercoordinated water molecules are referred to those with fewer than four nearest neighbors (CN < 4) as they occur in the bulk interior of water and ice. Molecular undercoordination occurs in the terminated O:H–O bonded networks, in the skin of a large or small volume of water and ice, molecular clusters, ultrathin films, snowflakes, clouds, fogs, nanodroplets, nanobubbles, and in the vapor phase. Such a kind of water molecules show dextrorotary feature of supersolidity such as the lowered freezing temperature $T_N$ [2-6] and raised melting point $T_m$, hydrophobic, less dense, superfluidity in microchannels [7]. Thin ice can form at room temperature on $SiO_2$ substrate [8]. The superfluidity between graphene sheets occurs only at a thickness of six layers or less [9].

Excessive properties [10-19] also include the longer O–O distance, H–O phonon blueshift and O:H phonon redshift. The O 1s binding energy shifts positively to stronger binding energy, nanobubbles are long lived, mechanical stronger and thermally more stable, skins of ice and water are hydrophobic and frictionless, moving faster in microchannels. Undercoordinated molecules have longer lifetime extending from the bulk value of 200 fs to some 700 fs. These features become more pronounced as the molecular coordination number decreases or the nanosolid curvature increases.

Salt solvation differs the local physical–chemical properties of hydrogen bonds in the hydration shells from those of the ordinary bulk water. Intensive pump–probe spectroscopic investigations have been conducted to pursue the mechanism behind molecular performance in the spatial and temporal domains. For instance, the sum frequency generation (SFG) spectroscopy resolves information on the molecular dipole orientation or the skin dielectrics, at the air–solution interface [20, 21], while the ultrafast two–dimensional infrared absorption probes the solute or water molecular diffusion dynamics in terms of phonon lifetime and the viscosity of the solutions [22, 23].

Salt solutions demonstrate the Hofmeister effect [24, 25] on regulating the solution surface stress and the solubility of proteins with possible mechanisms of structural maker and breaker [26-28], ionic specification [29], quantum dispersion [30], skin induction [31], quantum fluctuation [32], and solute–water interactions [33]. Increasing the chloride, bromide and iodide solute concentration shifts more the H–O stretching vibration mode to higher frequencies [34, 35]. These spectral changes are usually explained as the Cl$^-$, Br$^-$, and I$^-$ ions weakening of the surrounding O:H nonbond (structure breakers). An external electric field in the $10^9$ V/m order slows down water molecular motion and even crystallizes the system. The field generated by a Na$^+$ ion acts rather locally to reorient and even hydrolyze its neighboring water molecules according to MD computations [36].

However, knowledge insufficiency about O:H–O bond cooperativity [37] has hindered largely the progress in understanding the solvation bonding dynamics, solute capabilities, and inter– and intramolecular interactions in the salt solutions. One has been hardly able to resolve the network O:H–O bond segmental cooperative relaxation induced by salt solvation. It is yet to be known how the cation and anion interact with water molecules and their neighboring solutes, and their impact on the performance of the solutions such as the surface stress, solution viscosity, solution temperature, and critical pressures and temperatures for phase transition [34, 38]. Fine–resolution detection and consistently deep insight into the intra– and intermolecular interactions and their consequence on the solution properties have been an area of active study.

However, charge injection by salt solvation shares the same effect of molecular undercoordination on the phonon frequency shift [39-41]. One of the most appealing observations is that the melting of ice in porous glass having different distribution of pores sizes. The confined water crystallizes only partially and at an interface layer, between the ice crystallites and the surface of the pore, remains liquid. Nuclear magnetic resonance and differential scanning calorimetry measurements revealed a 0.5 nm thick interface [42].

Salt hydration and water confinement have been intensively investigated using the following multiscale approaches:
1) Classical continuum thermodynamics [43-47] embraced the dielectrics, diffusivity, surface stress, viscosity, latent heat, entropy, nucleation, and liquid/vapor phase transition in terms of free energy, though this approach has faced difficulties in dealing with solvation dynamics and the properties of water and ice.
2) Molecular motion dynamics (MD) [48-50] computations and the ultrafast phonon spectroscopies

are focused on the spatial and temporal performance of water and solute molecules as well as the proton and lone pair transportation behavior. Information includes the phonon relaxation or the molecular residing time at sites surrounding the solute or under different coordination conditions or perturbations.

3) Nuclear quantum interactions [51-53] simulation has enabled visualize the concerted quantum tunneling of protons within the water clusters and quantify the impact of zero–point motion on the strength of a single hydrogen bond at a water/solid interface. An interlay of STM/S and the *ab initio* path–integral molecular dynamics (PIMD) verify unambiguously that the $sp^3$–orbital hybridization takes place at 5 K temperature. The proton quantum interaction elongates the longer part and shortens the shorter part of the O:H–O bond.

4) Hydrogen bond (O:H–O) cooperativity [1, 54-56] enables resolution to multiple mysteries of water and ice. A combination of the Lagrangian mechanics, MD and DFT computations with the static phonon spectrometrics has enabled quantification of O:H–O transition from the mode of ordinary water to the conditioned states. Obtained information includes the fraction, stiffness, and fluctuation order transition upon perturbation and their consequence on the solution viscosity, surface stress, phase boundary dispersity, and the critical pressure and temperature for phase transition.

Experimental detection [57] and MD computations [58] showed consistently that both salt solvation and water confinement not only slow down MD dynamics characterized by the phonon lifetime but also shift the phonon frequency to higher frequencies. One needs to answer why do salt solvation and nanoconfinement transit the phonon lifetime and phonon stiffness in the same manner and what intrinsically dictates the chemical and physical properties of the confined water and the salt solution.

According to Pauling [59], the nature of the chemical bond bridges the structure and property of a crystal and molecule. Therefore, bond formation and relaxation and the associated energetics, localization, entrapment, and polarization of electrons mediate the macroscopic performance of substance accordingly [39]. O:H–O bond segmental disparity and O–O repulsivity form the soul dictating the extraordinary adaptivity, cooperativity, recoverability, and sensitivity of water and ice [54]. The proper answer to these questions is the chemical bond [59] and the valence electrons [39, 54]. One must focus on the bond relaxation and electron polarization in the skin region or in the hydration volume and interplaying of the afore–mentioned multiscale approaches is necessary.

We show here the electronic and phononic spectrometric evidence for the bond–electron–phonon correlation of the confined and hydrated hydrogen bond network and the supersolid state due to molecular undercoordination and electric polarization by salt solvation.

2. O:H−O Bond Oscillator Pair Scheme
2.1. Basic Rules for Water

Water prefers the statistic mean of the tetrahedrally–coordinated, two–phase structure in a core–shell fashion of the same geometry but different O:H−O bond lengths [1, 54]. Figure 1a illustrates the 2H$_2$O unit cell of C$_{3v}$ symmetry having four hydrogen bonds bridging oxygen anions. As the basic structure and energy exchange unit, the O:H−O bond integrates the intermolecular weaker O:H nonbond (or called van der Waals bond with ~0.1 eV energy) and the intramolecular stronger H−O polar–covalent bond (~4.0 eV) with asymmetrical and short–range interactions and coupled with the Coulomb repulsion between electron pairs on adjacent oxygen ions[1]. O:H−O length and bond angle relaxation changes system energy, but fluctuation contributes little to the system energy on average.

It is essential to treat water as a crystalline–like structure with well–defined lattice geometry, strong correlation, and high fluctuation. For a specimen containing N oxygen atoms, the 2N numbers of protons H$^+$ and lone pairs ":" and the O:H−O bond configuration conserve regardless of structural phase[60] unless excessive H$^+$ is introduced by acid solvation to form the H$_3$O$^+$ and H↔H anti–HB formation[61] or ":" introduction with HO$^-$ by base solvation to form the O:⇔:O super–HB [62]. The H$_3$O$^+$ or the HO$^-$ replaces the central H$_2$O in Figure 1a but the neighboring H$_2$O remain their orientations because of the lattice geometry and intermolecular interactions. The motion of a H$_2$O molecule or the proton H$^+$ is subject to restriction. If a molecule rotating above 60° around its C$_{3v}$ symmetrical axis, there will be H↔H anti–HB and O:⇔:O super–HB formation that is energetically forbidden. Because of the H−O bond energy of ~4.0 eV, translational tunneling of the H$^+$ is also forbidden. Breaking the H−O or the D−O bond in vapor phase requires 121.6 nm laser radiation [63, 64], estimated 5.1 eV because the extremely low molecular coordination numbers.

## 2.2. O:H–O Bond Potentials and Cooperativity

Figure 1b illustrates the asymmetrical, short–range, coupled three–body potentials for the segmented O:H−O bond [55, 56]. The proton serves as the coordination origin. The left–hand side is the O:H van der Walls (vdW) interaction and the right–hand side is the H−O polar covalent bond. The Columbo repulsion between electron pairs on neighboring $O^{2-}$ couple the O:H−O bond to be an oscillator pair.

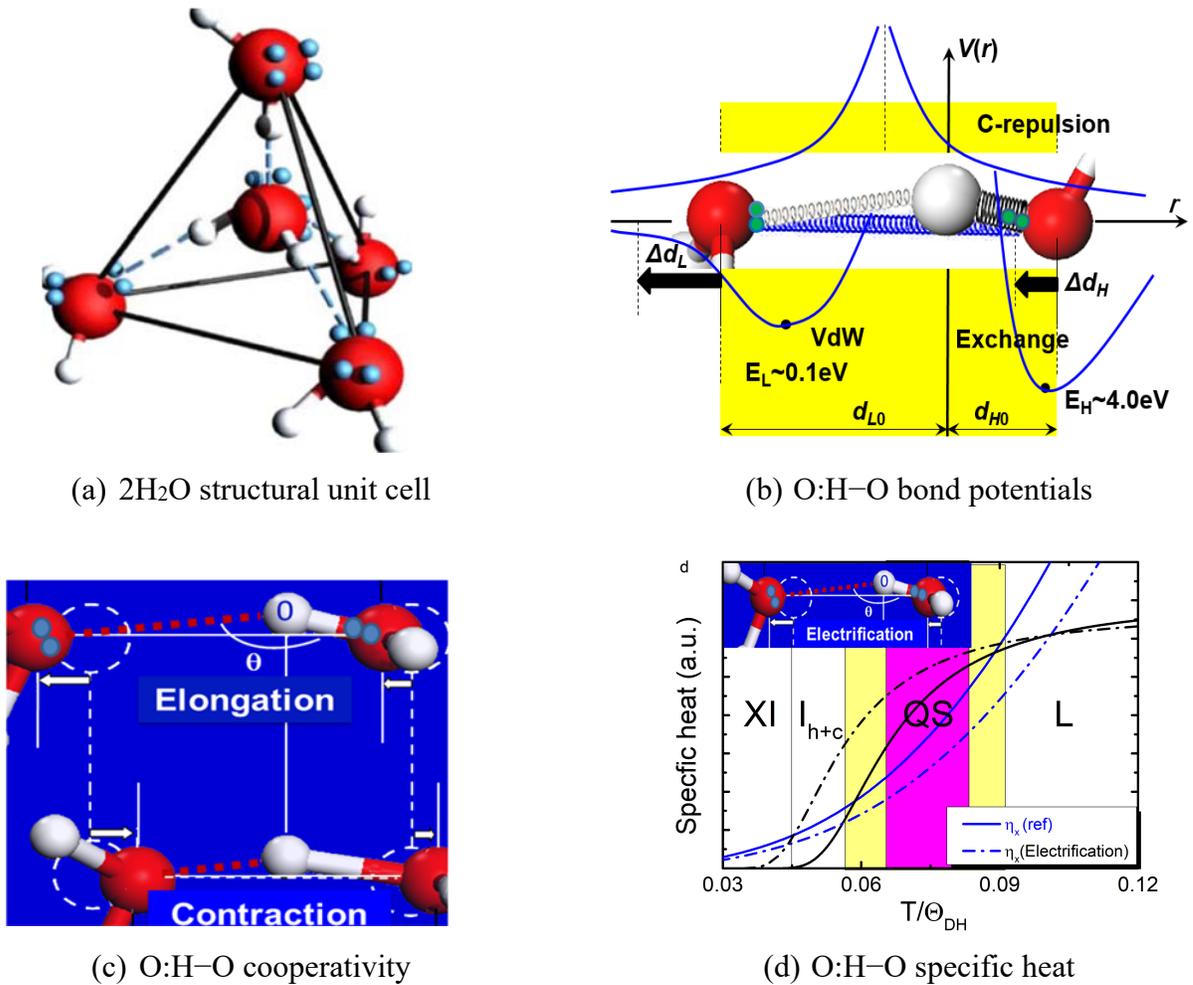

(a) 2H$_2$O structural unit cell

(b) O:H−O bond potentials

(c) O:H−O cooperativity

(d) O:H−O specific heat

Figure 1. (a) The 2H$_2$O primary unit cell contains four oriented O:H−O bonds defines liquid water as a crystal with molecular and proton motion restrictions[1]. (b) The asymmetrical, short–range, coupled three–body potentials for the segmented O:H−O bond[55, 56]. (c) Cooperative relaxation of the segmental O:H−O bond proceeded by elongating one part and contracting the other with respect to the $H^+$

coordination origin. The softer O:H always relaxes more than the stiffer H−O. The ∠O:H−O angle θ does not contribute to the bond length and energy. (d) Superposition of the segmental specific heat $\eta_x$ defines the phases from high temperature downward of the Vapor ($\eta_L = 0$, not shown), Liquid and ice $I_{h+c}$ ($\eta_L/\eta_H < 1$), quasisolid (QS) ($\eta_L/\eta_H > 1$), XI ($\eta_L \cong \eta_H \cong 0$), and the QS boundaries ($\eta_L/\eta_H = 1$) closing to $T_m$ and $T_N$, respectively[2]. Electrification (ionic polarization)[40] or molecular undercoordination [41] disperses the QS boundaries outwardly.

The O:H nonbond and the H−O bond segmental disparity and the O−O coupling allow the segmented O:H−O bond to relax oppositely – an external stimulus dislocates both O ions in the same direction but by different amounts, see Figure 1c. The softer O:H nonbond always relaxes more than the stiffer H−O bond with respect to the H$^+$ coordination origin. The ∠O:H−O angle θ relaxation contributes only to the geometry and mass density. The O:H−O bond bending has its specific vibration mode that does not interfere with the H−O and the O:H stretching vibrations [1]. The O:H−O bond cooperativity determines the properties of water and ice under external stimulus such as molecular undercoordination [5, 6, 65-67], mechanical compression [34, 38, 68-70], thermal excitation [2, 71, 72], solvation [73, 74] and determines the molecular behavior such as solute and water molecular thermal fluctuation, solute drift motion dynamics, or phonon relaxation (reprinted with permission from [2, 41, 75])

## 2.3. Specific Heat and Phase Transition

Figure 1d shows the superposition of the specific heat $\eta_x(T/\Theta_{Dx})$ of Debye approximation for the two segments (X = L and H for the O:H and H–O segment, respectively) [2]. The segmental specific heat meets two conditions. One is the Einstein relation $\omega_x \propto \Theta_{Dx}$ and the other is the thermal integration being proportional to bond energy $E_x$. The ($\omega_x$, $E_x$) is (200 cm$^{-1}$, ~0.1 eV) for the O:H nonbond and (3200 cm$^{-1}$, ~4.0 eV) for the H–O bond. The Debye temperatures and the specific heat curves are subject to the $\omega_x$ that varies with external perturbation. Thus, the superposition of segmental specific heat $\eta_x$ defines the phases from high temperature downward of Vapor (not shown), Liquid, Quasisolid (QS), $I_{h+c}$ ice, XI, and the QS boundaries of extreme densities and closing to $T_m$ and $T_N$, respectively [2].

The hydrogen bonding thermodynamics is subject to the specific heat ratio, $\eta_L/\eta_H$. The segmental having a lower specific heat follows the regular thermal expansion but the other segment responds to thermal excitation oppositely, which explains why ice floats when cooling at the QS phase – the H-O contracts less than O:H expansion at cooling. In the Vapor phase, $\eta_L \cong 0$, the O:H interaction is negligible; In the Liquid and $I_{c+h}$ ice, $\eta_L/\eta_H <1$, cooling contraction takes place at different rate; In the QS phase, $\eta_L/\eta_H > 1$, cooling expansion occurs; In the XI phase, $\eta_L \cong \eta_H \cong 0$, neither O:H nor H–O responds sensitively to temperature change [76, 77]; at the QS boundaries ($\eta_L/\eta_H = 1$) density transits with the boundaries closing to $T_m$ and $T_N$ [2].

3. Supersolidity and quasisolidity

The concept of supersolidity was initially extended from the $^4$He fragment at mK temperatures, demonstrating elastic, repulsive and frictionless between the contact motion of $^4$He segments [78] because e of atomic undercoordination induced local densification of charge and energy and the associated polarization [79]. The concepts of supersolidity and quasisolidity were firstly defined for water and ice in 2013 by Sun et al [2, 41] and then intensively verified.

The quasisolidity describes phase transition from Liquid density maximum of one gcm$^{-3}$ at 4 °C to Solid density minimum of 0.92 gcm$^{-3}$ at –15 °C, which demonstrates the cooling expansion because the specific heat ratio $\eta_L/\eta_H < 1$, the H–O bond contraction drives the O:H expansion and the ∠O:H–O angle relaxation from 160 to 165°.

The supersolidity features the behavior of water and ice under polarization by coordination number reduction or charge injection. When the nearest CN number is less than four the H–O bond contracts spontaneously associated with O:H elongation and strong polarization. At the surface, the H–O bond contracts from 1.00 to 0.95 Å and the O:H expands from 1.70 to1.95 Å associated with the O:H vibration frequency transiting from 200 to 75 cm$^{-1}$ and the H–O from 3200 to 3450 cm$^{-1}$ [80]. The shortened H–O bond shifts its vibration frequency to a lower value that increases with the reduction of the molecular CN.

The shortened H–O bond shifts its vibration frequency to a higher value that increases further with the reduction of the molecular CN, which disperses the QS boundaries outwardly, causing the supercooling at freezing and superheating at melting [15]. Under compression, the situation reverses, raising the $T_N$ and lowering $T_m$, which is the case of regelation – ice melts under compression and the $T_m$ reverse when the pressure is relieved [49]. The high thermal diffusivity of the supersolidity skin governs thermal transportation in the Mpemba paradox – warm water cools faster.

Salt solvation derives cations and anions dispersed in the solution [40]. Each of the ions serves as a source center of electric field that aligns, stretches and polarizes the O:H–O, resulting the same supersolidity in the hydration shell whose size is subject to the screening of the hydrating $H_2O$ dipoles and the ionic charge quantity and volume size.

Figure 2a and b show the ultrafast infrared spectral evidence for the supersolidity of water nanodroplets and NaBr solvation in a 5%$D_2O$ + 95%$H_2O$ solvent [57]. Replacing some H with D is for better resolution in detection. The pump–probe time–dependent IR spectroscopy probes the decay time of a known intramolecular (H–O or D–O) vibration phonon population decay or vibration energy dissipation lifetime through the solution viscosity and Stokes–Einstein relation for the molecular drift diffusivity [45], which is very much the same to optical fluorescent [81] and ultrafast photoelectron [82] spectroscopies. The signal lifetime is proportional in a way to the density and distribution of the defects and impurities. The impurity or defect states prevent the thermalization of the electron/phonon/photon transiting from the excited states to the ground for exciton (or electron–hole pair) recombination. One switches off/on the pump/probe simultaneously and monitors the population decay that features the rate of vibration energy dissipation during the wave propagation in the solution. Ultrafast IR spectroscopy revealed that the H−O phonon lifetime increases from 2.6 to 3.9 and 6.7 ps as the water transits into NaBr solution with concentration increasing from 32 to 8 $H_2O$ per NaBr and increases from 18 to 50 ps when the water droplet size is reduced from 4.0 to 1.7 nm [57].

The differential phonon spectrometrics (DPS) is the subtraction of the spectrum of bulk water from the those of the nanodroplets and salt solution upon all the spectral peak area being normalized. These DPS peaks show the transition of the D–O bond from the mode of ordinary water as valleys to the droplets and the hydrated mode, respectively, in terms of phonon stiffness ($\Delta\omega_D$), abundance (peak area), and fluctuation order (line width). The phonon peak frequency follows the $\omega^2 \propto E(\mu d^2)^{-1}$ relation with $\mu$ being reduced mass and $E$ the binding energy and $d$ the length of the H–O or D–O vibrating dimer.

Therefore, the phonon abundance transition from the valley to the peak in both cases show that the D–O bond turns to be shorter and stiffer in the hydration shell and in the droplet skin.

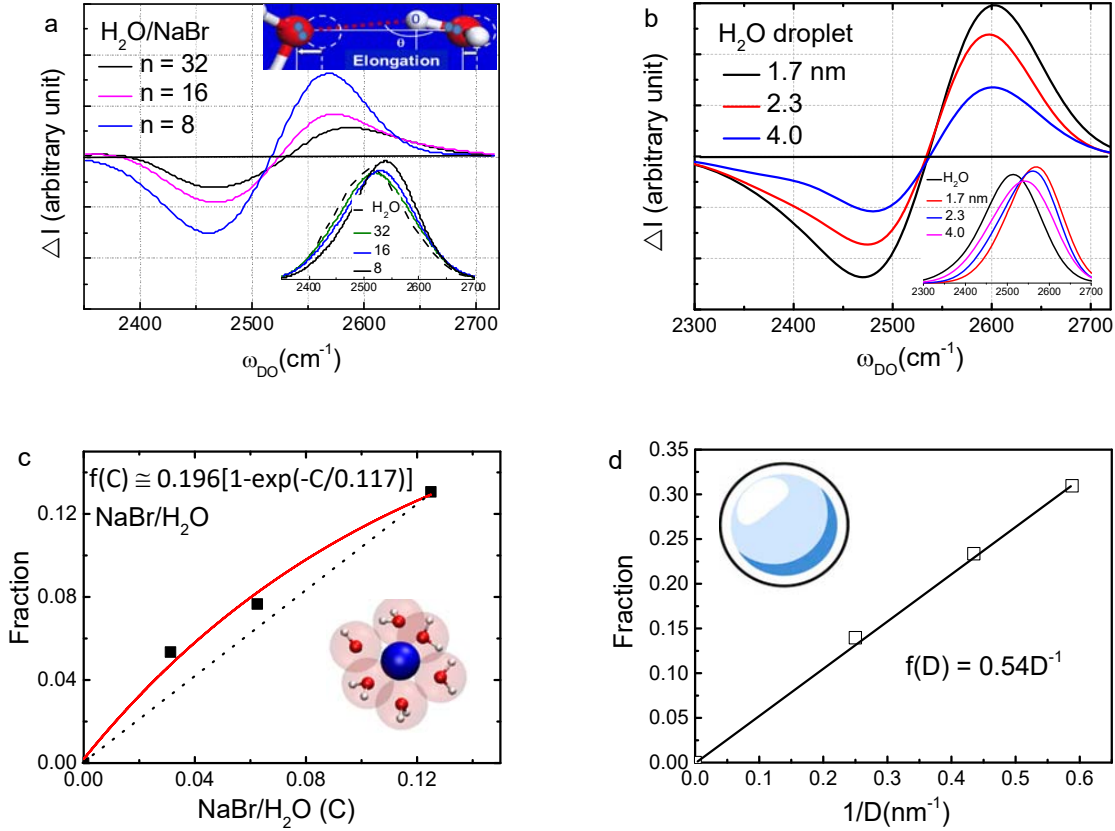

Figure 2. (a, b) D–O phonon DPS and (c, d) the fraction coefficient for (a, c) the concentrated NaBr solutions and (b, d) the sized water droplets. Insets a and b show the D–O peaks presented in [57]. Inset a illustrates the O:H–O elongation by polarization due to salt solvation and molecular undercoordination [40, 41].

An integration of the DPS peak gives rise to the fraction of bonds transiting from the mode of water to polarization in the ionic hydration volume and in the skin of the droplets, as shown in Figure 2c and d. The solute concentration $C$ and the droplet size $D$ resolved fraction coefficients and their slops follow the relations:

$$f_{NaBr}(C) \propto 0.196\,[1-\exp(-C/0.117)];\ df_{NaBr}(C)/dC = 1.675\exp(-C/0.117)$$

$f_{droplet}(D) = 0.54D^{-1}$; $df_{droplet}(D)/d(1/D) = 0.54$

The exponential drop of the slope $df_{NaBr}(C)/dC$ indicates the hydration shell volume or the solute electric filed decreases with solution density, arising from anion− anion repulsion [40]. The small cations have been fully screened by the hydrating dipoles and thus cations do not interact with any other solute. The $df_{NaBr}(C)/dC$ clarifies the presence of the supersolid covering sheet of a constant thickness $\Delta R$. One can estimate the shell thickness $\Delta R$ of a spherical droplet of $V \propto R^3$: $f(R) \propto \Delta V/V = \Delta N/N = 3\Delta R/R$, and the skin thickness $\Delta R_{skin} = f(R)R/3 = 0.90$ Å is exactly the dangling H–O bond length [1] featured at 3610 cm$^{-1}$ wavenumber of vibration [41]. The next shortest H–O bond in the skin is about 0.95 Å featured at 3550 cm$^{-1}$ [83]. The DPS distills only the first hydration shell and the outermost layer of a surface [84] without discriminating the intermediate region between the bulk and the outermost layer.

The $\Delta\omega_H > 0$ and $\Delta\omega_L < 0$ disperse the Debye temperature $\Theta_{Dx}$ accordingly and hence disperses the QS boundaries outwardly, as illustrated Figure 1d, leading to the supercooling at freezing and superheating at melt, as one observes as the "no man's land" temperature as shown in Figure 3a and the freezing temperature depression by salt solvation. XRD, Raman and MD observations show that 1.2 nm sized droplet freezes at 173 K [5], and the (H$_2$O)$_{3-18}$ clusters do not form ice even at 120 K [6]. The XI-I$_c$ phase transition temperature also varies with the droplet size [76, 77].

From the $f(1/D)$ in Figure 2d and the fact of the droplet sized resolved $T_N$, one can infer that a droplet holds the two phase structure in terms of coordination− resolved core− shell configuration rather than the domain resolved high/low density random patches. Figure 3b shows the salt concentration resolved $T_N$ depression of salt solutions.

Because of the strong polarization and $\Delta\omega_L < 0$, the surface of water and ice is offered with a supersolid skin that is elastoviscous, less dense (0.75 unit), and mechanically and thermally more stable. The high elasticity and soft O:H phonon adaptivity and the densely packed dipoles ensure the slipperiness of ice [85] the nanobubble endurability [86] and toughen the water skin [87]. It is the supersolid skin lower density that ensure the high thermal diffusivity for the heat transport in the Mpemba effect [88].

One must note that the QS boundaries are not constant but change with the phonon frequency relaxation under external excitation. Electrification (ionic polarization) [40] or molecular undercoordination [41] shift the H–O phonon frequency up and the O:H frequency down, which disperses the QS boundaries outwardly, resulting in the freezing temperature $T_N$ depression and melting temperature elevation [40, 41]. However, compression disperses the boundaries inwardly, lowering the melting temperature and raising the freezing temperature, call Regelation [89].

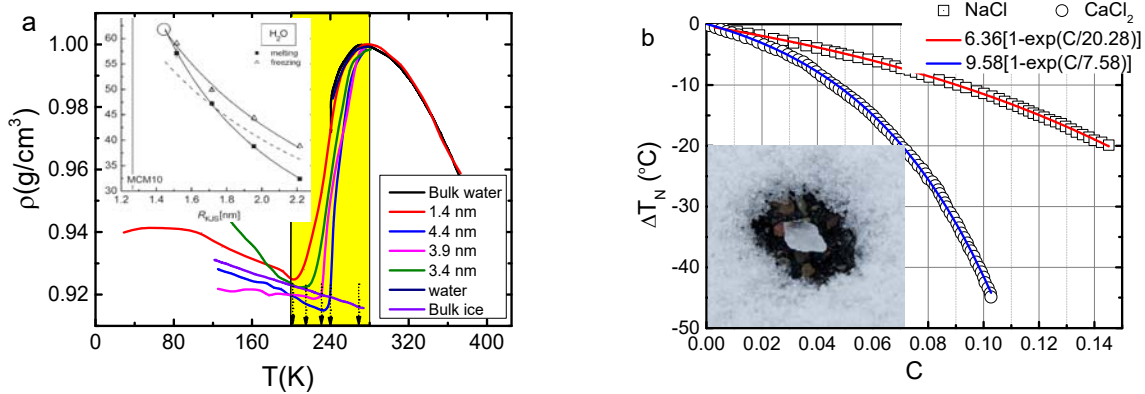

Figure 3. (a) $H_2O$ droplet size resolved $T_N$ [2-6] and (b) $T_N$ depression by concentrated NaCl and $CaCl_2$ solvation [90] with inset showing the airport salt anti–icing. (c) molecular undercoordination dispersed QS boundary [91]. Inset a is reprinted from [92] for the Tc shift. (c) Mechanism of the quasisolid (QS) phase boundary dispersion by salt solvation, which raises the $T_m$ and depresses the $T_N$ of the solution. (Reprinted with permission from [2, 90, 93]).

4. Electron and Phonon Spectrometrics

4.1. STM and STS: Strong Polarization

Figure 4 shows the orbital images and the dI/dV spectra of a $H_2O$ monomer and a $(H_2O)_4$ tetramer deposited on a NaCl(001) surface probed using STM/S at 5 K [94]. The highest occupied molecular orbital (HOMO) below the $E_F$ of the monomer appears as a double−lobe structure with a nodal plane in between, while the lowest unoccupied molecular orbital (LUMO) above $E_F$ appears as an ovate lobe developing between the two HOMO lobes. STS spectra at different depths discriminate the tetramer from the monomer in the density of states (DOS) crossing $E_F$.

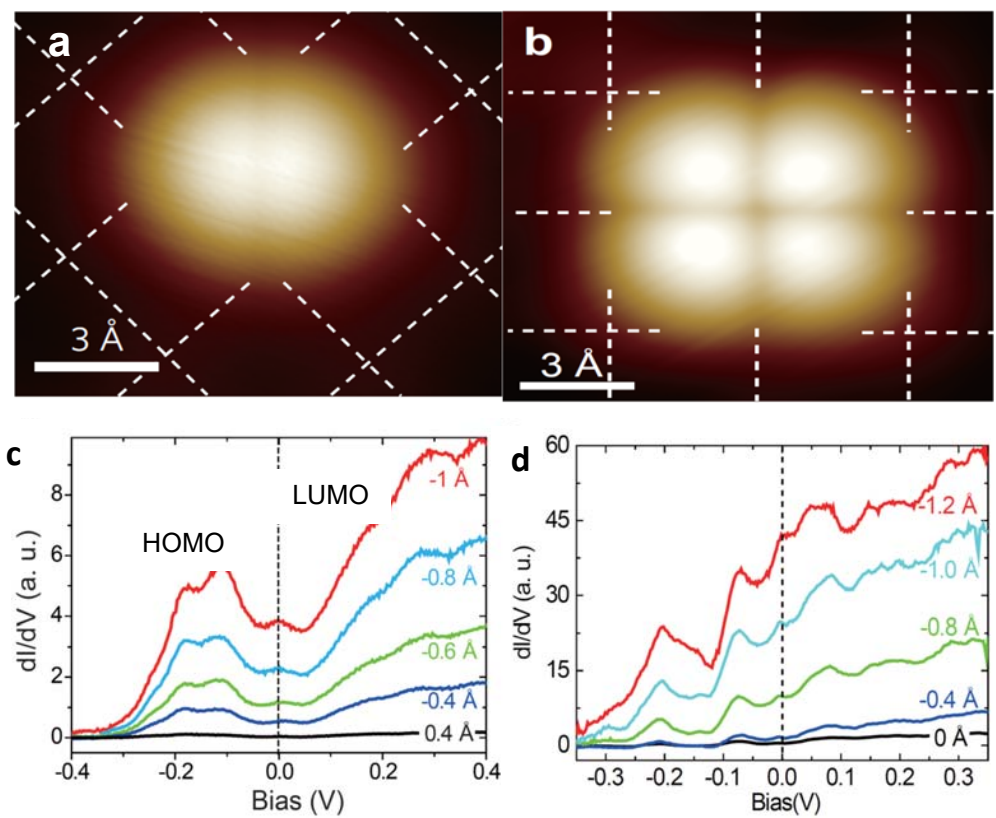

Figure 4. STM images of (a) a $H_2O$ monomer and (b) a $(H_2O)_4$ tetramer, and (c, d) the respective dI/dV spectra obtained under conditions of V = 100 mV, I = 100 pA, and dI/dV collected at 50 pA of different heights at 5 K temperature. Grid in images denotes the $Cl^-$ lattice of the NaCl(001) substrate (Reprinted with permission from [94].) LUMO (> $E_F$) and the HOMO (< $E_F$) indicated in (b) denote the orbital energy states.

These STM images [94] confirmed the occurrence of $sp^3$−orbit hybridization of oxygen in $H_2O$ monomer occurs at 5 K or lower and the intermolecular interaction involved in $(H_2O)_4$. Therefore, the 2N number conservation holds regardless of temperature. Even at extreme conditions of 2000 K temperature and 2 TPa pressure, $2H_2O$ transits into $H_3O^+$:$HO^-$ [95], the 2N number conservation of protons and lone pairs remains. According to the bond−band−barrier correlation [39, 96], the HOMO located below $E_F$ corresponds to the energy states occupied by electron lone pairs of oxygen, and the LUMO to states yet to be occupied by electrons of antibonding dipoles. The image of the monomer showing the directional lone pairs suggests that the lone pairs point into the open end of the surface. As the $H^+$ ion can only share its unpaired electron with oxygen, the Cl⁻ ion in the NaCl substrate interacts with the H⁺ only electrostatically.

## 4.2. Molecular Site Resolved H−O Bond Characteristics

The sum frequency generation (SFG) probe the molecular orientation and dielectrics at the interface as a function of vibration frequency [20]. SFG measurements, shown in Figure 5a [97], confirmed that in the outermost subsurface between the first (B1) and the second (B2) bilayer, the O:H nonbond of the $O_{B1}$−H:$O_{B2}$ is weaker than that of the $O_{B1}$:H−$O_{B2}$. The subsurface O−O distance is laterally altered, depending on the direction of H−O bond along the surface normal: H–up or H–down, which is in contrast with bulk O:H nonbonds. This discovery is consistent with the DFT derivatives [98] and the present O:H−O bond cooperativity notion expectation [55, 56]. The least coordinated outermost sublayered H−$O_{B1}$ is shorter and stiffener than the second sublayered H−$O_{B2}$ and hence the $O_{B2}$:H is longer and softer than the $O_{B1}$:H.

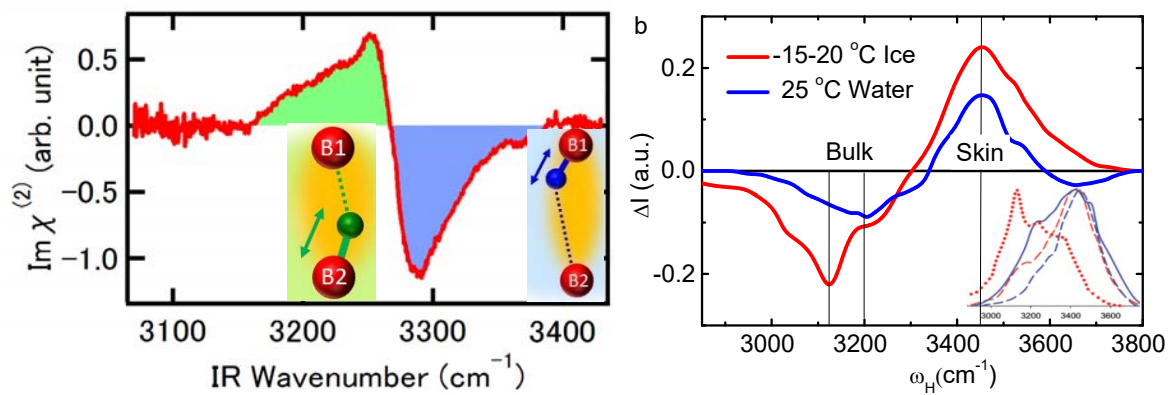

Figure 5. SFG spectrum for the H–O oscillators lined along the c–axis of ice Ih(0001). Insets illustrate the $O_{B1}$:H−$O_{B2}$ and the $O_{B1}$−H:$O_{B2}$ segmental bond lengths, orientations, and the H–O stretching vibration frequencies. The positive peak (< 3270 cm⁻¹) corresponds to the H−$O_{B2}$ (shaded in green) and the valley (> 3270 cm⁻¹) to the $O_{B1}$−H (shaded in blue). The less coordinated $O_{B1}$−H is shorter and stiffer and its H:$O_{B2}$ is longer and softer than the inverse (reprinted with permission from [97]). (b) the DPS of water and ice collected at 87 and 0° polar angles (inset b) [99] decomposed the $\omega_H$ into the bulk (3200 cm⁻¹) and skin (3450 cm⁻¹) components (inset a) with an additional feature for the H−O free radical at 3600 cm⁻¹

(reprinted with permission from [80])

The DPS profiles are obtained by subtracting the spectra collect at angle closing to the surface normal from the ones collected in gracing angles and all of them being spectra area normalized [99]. Figure 5b shows that the –(15− 20) °C ice and the 25 °C water share the same skin supersolidity – identically shortened and stiffened H–O bond.

DFT calculations by Wang and collaborators [47, 83] examined the site–resolved electronic binding energy and H−O stretching vibration frequency for medium sized $(H_2O)_n$ clusters (n = 17-25). They classify the H−O bonds into five groups: the dangling H−O bonds, the H−O bonds associated with the dangling $H_2O$ molecules, those with undercoordinated $H_2O$ molecules without dangling H−O bonds, those forming the tetrahedral–coordination of the central $H_2O$ and the others in its neighbor four molecules. The neighboring four molecules and the outer undercoordinated molecules form cages covering the central $H_2O$. Computations resolved that these two regions interact competitively, showing complementary interaction energy with the change of cluster size. Raman spectra in Figure 6 reveal the site–resolved vibration frequencies. The dangling bond frequency features keep constant while others vary with cluster size. Observations confirm the effect of coordination environments on the H−O bond contraction, H−O bond energy gain, and its vibration frequency blue shift [1].

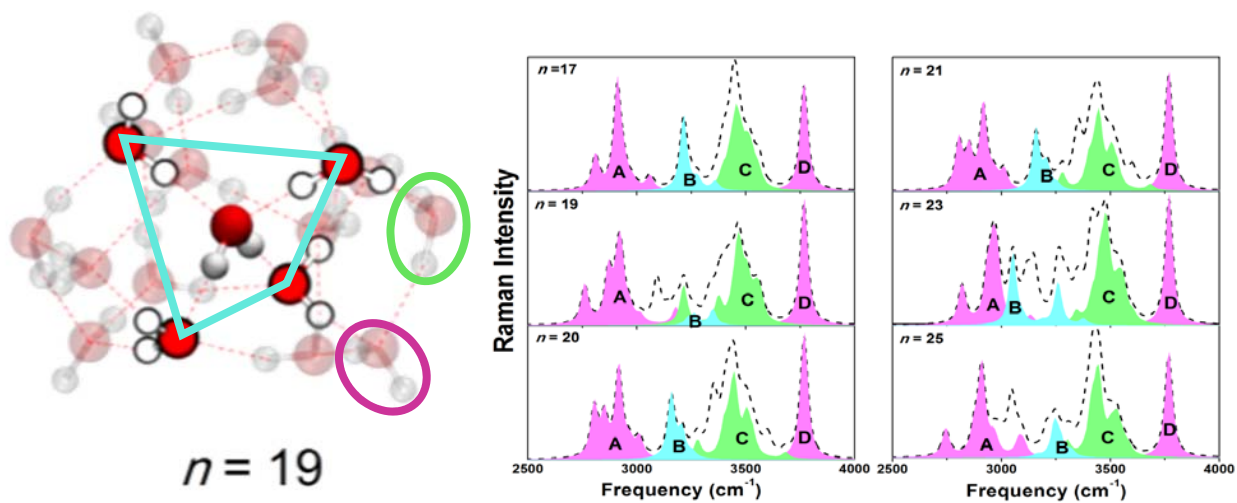

Figure 6. Computational H−O stretching vibration modes in the $(H_2O)_n$ clusters. The black dashed lines convolute the H−O vibration modes of the entire clusters. The sharp feature D corresponds to the H−O

dangling bonds, C to the H−O bonds associated to the undercoordinated molecules. Features A and B to the H−O bonds inside the clusters. Reprinted with copyright permission from [83].

4.3. Ultrafast PES and DPS: Nonbonding Electron Polarization

Molecular undercoordination induced skin polarization have been detected using an ultrafast pump− probe liquid−jet UPS [82]. The vertical bound energies (being equivalent to work function) of the hydrated electrons is 1.6 eV in the skin and 3.2 eV in the bulk interior for the hydrated electrons in pure water. The bound energy decreases with the number $n$ of the $(H_2O)_n$ clusters toward zero [100-102]. The hydrated electrons live longer than 100 ps near the surface compared with those solvated inside liquid bulk interior. Observations evidence that molecular undercoordination substantially enhances nonbonding electron polarization [41], which increases the viscoelasticity and hence lowers the skin molecular mobility. The anchored skin dipoles allow nanodroplet interacting with other substance through electrostatic, van der Waals, and hydrophobic interactions without exchanging electrons or bond formation, named non-additivity [103].

The nonbonding electrons are subject to dual polarization when the molecular CN is reduced [1]. Firstly, H−O bond contraction deepens the H−O potential well and entraps and densifies electrons in the H−O bond and those in the core orbitals of oxygen. This locally and densely entrapped electrons polarize the lone pair of oxygen from the net charge of –0.616 e to –0.652 eV accoring to DFT calculation for ice skin [80]. The increased charge of O ions further enhances the O−O repulsion as the second round of polarization. This dual polarization rasies the valence band energy up.

Further reduction of cluster size, or the molecular CN, enhances this dual polarization, resulting observations in Figure 7a – cluster trend of the solvate electron polarization. Therefore, electron dipoles formed on the flat and the curved skins enhaces such polarization, which creates the repulsive force, making liquid water hydrophobic and ice slippery. Nonbonding electron polarization notion indicates that molecular undercoordination polarizes nonbonding electrons in two rounds by the densely entrapped H−O bonding electrons and by then the repulsion between electron pairs on adjacent oxygen anions [1].

Hydrated electrons provide a probe for the cluster size and molecular site resolved information on their bound energy and lifetime. Using the ultrafast pump–probe photoelectron spocpy, Verlet et al [14] discovered that an excess electron can bound to the surface of a water cluster and to the ambient water/air interface. The internally solvated electron bound energy for $(D_2O)_{50}^-$ is centered at –1.75 eV and surface localized states are centered at –0.90 eV. These two states vary with the cluster size and from $(D_2O)_{50}^-$ to $(H_2O)_{50}^-$ slightly.

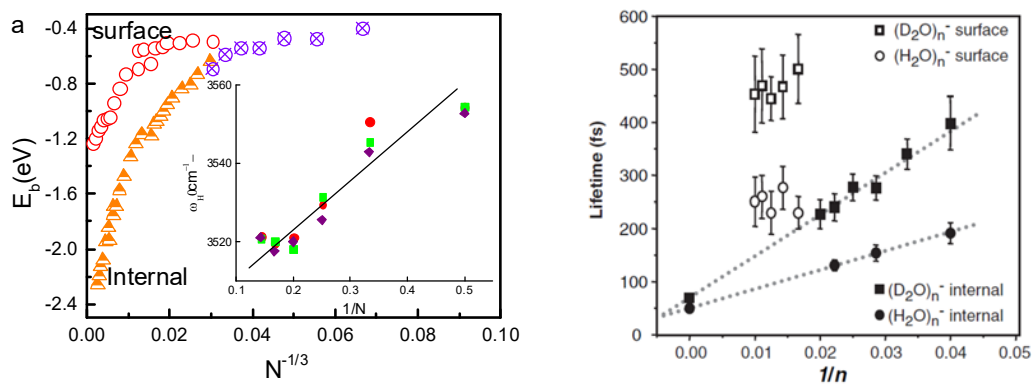

Figure 7. (a) $(H_2O)_n^-$ (n = 2-11) cluster size and molecular site resolved bound energy of hydrated electron [100-102] with inset showing the size resolved H−O phonon frequency [104-106]; (b) The lifetime of hydrated electrons [14]). The relaxation time scales are given after electronic excitation of the surface isomers at 0.75 eV (open symbols) and internal isomers at 1.0 eV (solid symbols) of $(H_2O)_n^-$ (circles) and $(D_2O)_n^-$ (squares) (Reprinted with permission from [82].)

Figure 7a shows that the hydrated electron can reside at the water/air interface, but remain below the dividing surface, within the first nanometer [100-102, 107]. Comparatively, the neutral $(H_2O)_N$ size–resolved H−O vibration frequency (inset a) and photoelectron lifetime (Figure 7b) shifts linearly with the inverse of cluster size [104-106, 108]. The lifetime of the surface states is much longer than that of the internal states. These cluster size and molecular site resolved electron bound energy, phonon stiffness, and electron phonon lifetimes confirm consistently the molecular undercoordination induced supersolidity.

The H−O phonon lifetime is proportional to frequency [23]. The polarization not only shortens the H−O bond and stiffens its phonons but also lengthens the O:H nonbond and softens the O:H phonon. The polarization lowers the bound energy of the hydrated electrons. The extent of polarization increases with decreases of cluster size and the surface polarization is more significant than it is at the interior of the cluster. The invariant photoelectron lifetime at the water cluster surface coincides with the extraordinary thermal stability of the supersolid skin [109].

4.4.  XPS and XAS: O 1s Energy Shift

Following the same trend as "normal" materials, molecular undercoordination imparts to water local charge densification [14, 15, 82, 110-112], binding energy entrapment, and nonbonding electron polarization [82]. Figure 8a shows that the O1s level shifts more deeply from the bulk value of 536.6 eV to 538.1 eV and to 539.7 eV when one moves from the bulk interior to its skin and then to a monomer in gaseous phase [113-115]. The O1s binding energy shift is a direct measure of the H−O bond energy, $\Delta E_{1s} \propto E_H$, and the contribution from the O:H nonbond is negligibly small, according to the Tight−Binding approximation [84].

Figure 8 b and c compares the NEXFAS spectra of nanobubbles [116], vapor, liquid skin, and bulk water [117]. The spectra show three majors at 535.0 to 536.8 and 540.9 eV corresponding, respetively, to the molecular coordination resolved bulk interior, skin, and H−O dangling bond radicals.

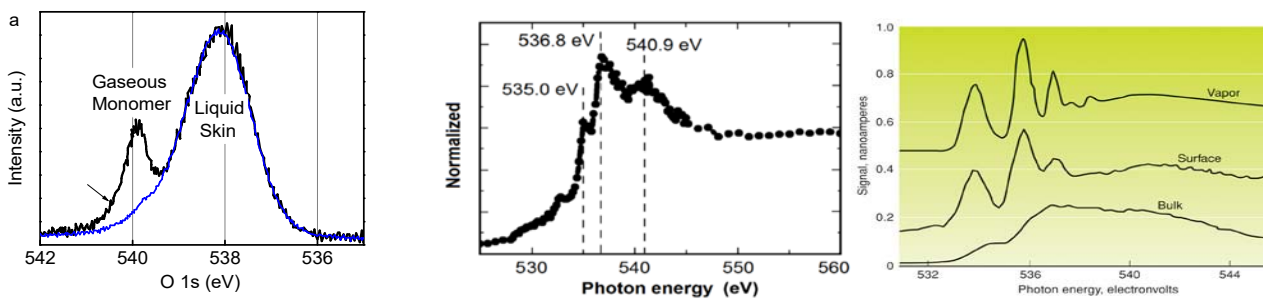

Figure 8. (a) XPS O 1s spectra of water containing emissions from the liquid skin at 538.1 eV and from the gaseous phase at 539.9 eV (reprinted with permission from [115]). (Reprinted with permission from [118].) NEXFAS spectra of (b) nanobubbles [116] and (c) vapor, liquid skin, and bulk liquid [117]. The spectra resolve discrete peaks that correspond to the bulk, skin, and H−O dangling bond radicals.

Comparatively, the energy conservation mechanism in the near edge X−ray fine structure absorption (NEXFAS) measurements is different from that of the XPS. The NEXFAS involves the shift of both the valence and the O 1s core band but the XPS involves the O 1s only. The NEXFAS pre–edge shift is the relative shift of the O1*s* core level $\Delta E_{1s}$ and its valence band shift $\Delta E_{vb}$ (occupied $4a_1$ orbital) from their energy levels of the isolated O atom: $\Delta E_{edge} = \Delta E_{vb} - \Delta E_{1s} \propto \Delta E_H > 0$ (H–O bond energy) [41]. The $\Delta E_{vb}$ is always greater than the $\Delta E_{1s}$ because of the shielding of electron in outer orbitals [84].

4.5. XAS and DPS: Supersolid H–O Bond Thermal Stability

Strikingly, the NEXFAS measurements [119] revealed that $Li^+$, $Na^+$, and $K^+$ cations shift the pre–edge component peak energy more than the $Cl^-$, $Br^-$, and $I^-$ anions and the pre–edge component associated with the first hydration shell of $Li^+$ ion is thermally more stable than those beyond, see Figure 9, as compared with the $\omega_H(T)$ thermal stability at site of the skin, bulk, and the dangling H–O bond of deionized water.

At 25 °C, the cation effect on the pre–edge shift from the value of 534.67 eV in alkali chlorides is remarkable: $Li^+$(0.27 eV), $Na^+$ (0.09 eV), and $K^+$(0.00 eV). The energy shift of $Li^+$ ion in 5 M LiCl solution (0.30 eV) is close to that in 3 M. On the other hand, in sodium halides, the anion effect is small: $Cl^-$(0.09 eV), $Br^-$(0.04 eV), and $I^-$ (0.02 eV). The energy trend of the pre–edge shifts is the same as the DPS $\Delta\omega_H$ of the solutions and the skin of water [109].

Table 1 lists the MD estimation and neutron diffraction resolved the first $O^{2-}:Y^+$ and $O^{2-}\leftrightarrow X^-$ hydration shell sizes of the solutes. MD calculations agree with the DFT derived segmental strains of the O:H–O bonds surrounding $X^-$ solutes [61].

The contributions from the Li:O polarization or the O:H binding energy change are negligibly small. At polarization, the H–O bond becomes shorter and stiffer, the stiffer H–O bond is thermally more stable than those in the bulk of ordinary water, as it does in the water skin [109]. The identical $O_{fw}$–$O_{bw}$ and $O_{bw}$–$O_{bw}$ thermal expansion in both the pure water and in the 5M LiCl solution indicates the invariance of the $Li^+$ hydration shell size, which does not interfere the O–O thermal behavior between the hydrated and non–hydrated oxygen anions.

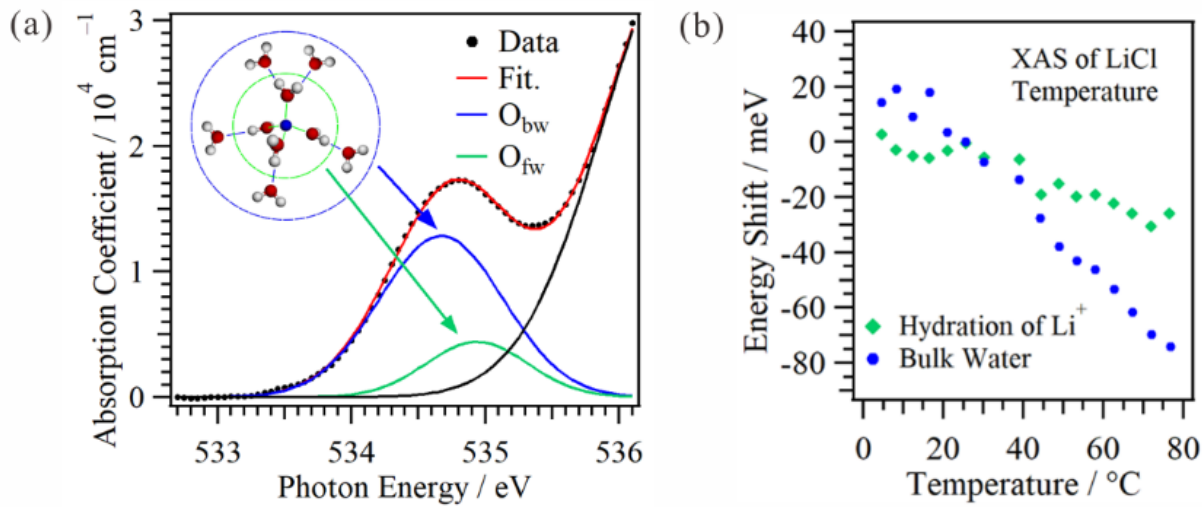

Figure 9. Site and temperature resolved (a) NEXFAS pre–edge component energy shift for LiCl solution. (b) The $E_{edge}$ for the first hydration shell ($O_{fw}$) shifts more than those beyond ($O_{bw}$) yet the $O_{fw}$ energy is thermally more stable. The energy shifts still undergo thermal entrapment but the hydration $E_{edge}$ shifts slower than that of pure bulk water (Reprinted with permission from [119]).

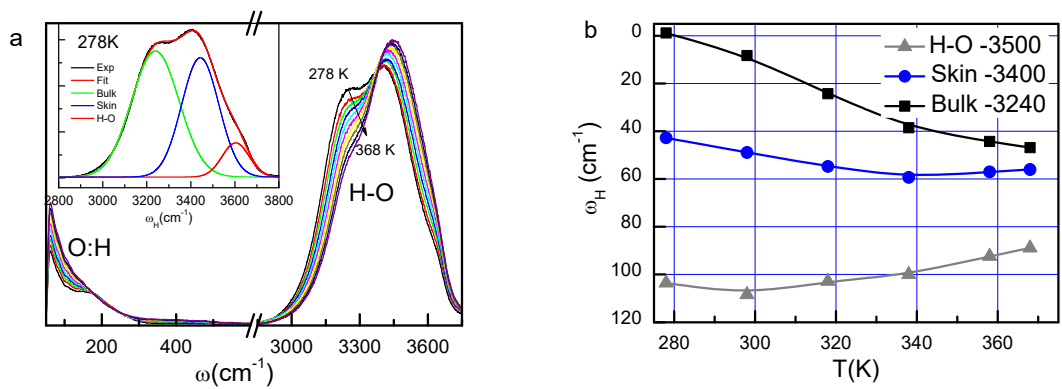

Figure 10. (a) Full–frequency Raman spectroscopy of deionized water heated from 5 to 95°C revealed that heating shifts the $\omega_H$ and the $\omega_L$ phonon frequencies in opposite detraction. The $\omega_H$ is decomposed into the bulk (3200 cm$^{-1}$) and skin (3450 cm$^{-1}$) components with an additional feature for the H−O free radical at 3600 cm$^{-1}$ (inset a). (b) The DPS skin $\omega_H(T)$ components is thermally more stable than the bulk showing both thermal H-O contraction but the dangling H-O bond thermal expansion because lacking the O-O coupling at the open end of water surface (Reprinted with permission from [109]).

The consistency between DPS of pure water and NEXFAS for salt solution evidences for the thermal stability of the supersolid states of the $Y^+$ hydration shell and the skin of pure water. The NEXFAS and DPS observations verified the following concepts:

1) The H–O bond energy undergoes thermal gain but for regular substance the bond energy is subject to thermal loss [120]. The H–O energy gain results from the O:H–O cooperativity when the O:H is elongated by ionic polarization, liquid heating [35], skin molecular undercoordination [41]. These observations showed that the O:H–O bond in the supersolid phase follows the regular thermodynamics outside the QS phase temperatures: $dd_L/dT > 0$, $dd_H/dT < 0$.

2) The shortened H–O bonds are thermally and mechanically more stable because the stiffened bonds are less sensitive to perturbation. The shorter H–O bond is harder to be further shortened by heating or to be lengthened by compression because of the O:H–O cooperativity nature [34, 38]: $(d|E_H|/dT)_{super}/(d|E_H|/dT)_{regular} < 1$ and $(d\omega_H/dT)_{super}/(d\omega_H/dT)_{regular} < 1$.

3) The QS upper boundary is at 4 °C for regular water and it seems at 25 °C for the supersolid states according to the curve slopes in Figure 9b.

4) The local electric field of small cations is stronger than that of a larger anions because of the $X^-\leftrightarrow X^-$ repulsion due to the high–order molecular structure of the supersolid hydration shells with insufficient number of dipoles screening the electric field [40].

Table 1. The $O:Y^+$ and $O^{2-}\leftrightarrow X^-$ distances (Å) in the YX solutions [119, 121, 122]

|  | $Li^+:O^{2-}$ (5M LiCl) | $Na^+:O^{2-}$ (Å) | $K^+:O^{2-}$ (Å) |
|---|---|---|---|
| $Y^+:O^{2-}$ distance MD [119] | 2.00 (5°C) | – | – |
|  | 1.99 (25°C) | 2.37 | 2.69 |
|  | 1.98 (80 °C) | – | – |
| Neutron diffraction [119, 121, 122] | 1.90 | 2.34 | 2.65 |
| MD [119] $X^-\leftrightarrow O^{2-}$ distance | $Cl^-\cdot H^+\text{–}O^{2-}$ | $Br^-\cdot H^+\text{–}O^{2-}$ | $I^-\cdot H^+\text{–}O^{2-}$ |
|  | 3.26 | 3.30 | 3.58 |
| DFT(acid) [61]* | $Cl^-\cdot(H\text{–}O:H)$ | $Br^-\cdot(H\text{–}O:H)$ | $I^-\cdot(H\text{–}O:H)$ |
| $1^{st}$ $(\varepsilon_H; \varepsilon_L)\%$ | –0.96; +26.1 | –1.06; +30.8 | –1.10; +41.6 |
| 2$^{nd}$ ($\epsilon_H$; $\epsilon_L$)% | –0.73; +19.8 | –0.78; +22.8 | –0.83; +28.6 |
| XAS (5 M LiCl) [119] | Li$^+$:(first O:H–O) | Li$^+$: (next O:H–O) | O:H–O (H$_2$O) |
| 5°C | $d_{O-O}$ = 2.71 (at 4 °C, $d_{O-O}$ = $d_H$ + $d_L$ = 1.0004 +1.6946 = 2.695 [75]) | | |
| 80 °C | 2.76 ($\Delta d_L$ > –$\Delta d_H$(<0)) | | |

*($\epsilon_H$; $\epsilon_L$)% is the DFT derived segmental strain for the first and the second O:H–O bonds radially surrounding X$^-$ anions in acid solutions. The strain is refereed the standard values of $d_H$ = 1.0004 and $d_L$ = 1.6946 Å for 4 °C water [75]. X$^-$·H represents the anions and H$^+$ Coulomb interaction.

## 5. Perspectives

A combination of the STM/S, XPS, XAS, ultrafast UPS, SFG, DPS, and ultrafast FTIR observations and quantum theory calculations reveals consistently the bond− electron− phonon correlation of the supersolidity of the confined and the hydrating water. Molecular undercoordination and charge dispersion by salt solvation share the same effect on the O:H− O relaxation and nonbonding electron polarization, which modulate the local hydrogen bonding network and the water properties through H−O bond contraction and O:H nonbond elongation. The O 1s energy shifts deeper, electronic vertical bound energy decreases but the electron and phonon lifetime become longer. The supersolid phase is less dense, elastoviscous, mechanically and thermally more stable with the hydrophobic and frictionless surface. The O:H−O bond cooperative relaxation disperses outwardly the quasisolid phase boundary to raise the melting point and meanwhile lower the freezing point of water ice.

From what we have experienced in the presently described exercises, one can be recommended the following ways of thinking and strategies of approaching for efficient investigation, and complementing to conventional approaches:

1) The nonbonding electron lone pairs pertained to N, O, F and their neighboring elements in the Periodic Table form the primary element being key to our life, which should receive deserved attention. It is related to DNA folding and unfolding, regulating, and messaging. NO medication and CF$_4$ anticoagulation in synthetic blood are realized through lone pairs interaction with living cells. The lone pair forms the O:H and the O:⇔:O interactions together with the H↔H determines the molecular interactions. However, the presence and functionality of the localized and weak lone pairs interactions have been oversighted. Without lone pairs, neither O:H–O bond nor oxidation could be possible; molecular interaction equilibrium could not be realized. Extending the knowledge about lone pairs and their functionality of polarization to catalysis, solution-protein, drug-cell, liquid-solid, colloid-matrix, interactions and even energetic explosives and other molecular crystals would be even more fascinating.

2) The key to the O:H–O bond is the O–O coupling. Without such a coupling none of the cooperative relaxation or the mysterious of water ice and aqueous solutions such as ice floating, ice slipperiness, regelation, supercooling/heating or the negative thermal expansion, warm water cooling faster. Unfortunately, the O–O coupling has been long overlooked in practice. Extending to the general situation containing lone pairs and X:A–Y bond, the impact would be tremendously propounding to the universe where we are living. For instance, a combination of O:⇔:O or other forms of repulsion and O:H–O(N) elongation could shorten the intramolecular covalent bonds for energy storage in the energetic materials such as full-N and N-based explosives because O:H–O bond elongation stiffens the H–O bond. Negative thermal expansion arises from segmental specific heat disparity of the X:H–Y and the superposition of the specific heat curves, such as graphite with discrepant inter- and intra-layer interactions.

3) It is necessary to think about water and solvent matrix as the highly ordered, strongly correlated, and fluctuating crystals, particularly, the supersolid phase caused by hydration and molecular undercoordination rather than the amorphous or multiphase structures. Water holds the two-phase structure in the core shell configuration, rather than the randomly domain-resolved mixture of density patches. Liquid water and the matrix of aqueous solutions must follow the conservation rules for the 2N number of protons and lone pairs and for the O:H–O configuration despite its segmental length and energy relaxation unless excessive H$^+$ or lone pairs are introduced.

4) One can consider the solvation as a process of charge injection with multiple interactions. Charge injection in the form of hydrated electrons, protons, lone pairs, cations, anions, molecular dipoles mediate the HB network and properties of a solution. Protons and lone pairs cannot stay alone but they are attached to H$_2$O molecule to form the H$_3$O$^+$ or HO$^-$ tetrahedron, respectively, breaking the conservation rule of water ice. Drift or Brownian motion of H$_3$O$^+$ or HO$^-$ may happen under electric or thermal fluctuation. It would be comprehensive to consider the electrostatic polarization and hydrating H$_2$O molecular screening, O:H, H↔H, O:⇔:O, solute-solute interactions and their variation with solute type and concentration.

5) Multifield perturbation supplies the basic degrees of freedom that mediate the solution HB network and properties by relaxing and transiting the O:H–O bond and electrons. H–O bond relaxation exchanges energy while the O:H relaxation or molecular motion dissipates energy capped at its cohesive energy about 0.01 eV.

6) The concepts of quasisolid (quasiliquid) of negative thermal expensity due to O:H–O bond segmental specific disparity, supersolidity due to molecular undercoordination and electric polarization, the H↔H anti-HB due to excessive protons, and the O:⇔:O super-HB due to excessive lone pairs are essential to describe the multifield effect on the performance of water ice and aqueous solutions. The quasisolid phase boundary dispersion by perturbation determines the solution O:H–O bond network and thermodynamic behavior.

7) A combination of the multiscale theory such as classical thermodynamics, DFT ad MD quantum computations, Nuclear quantum effect, and the O:H–O cooperativity and polarization notion to overcome limitation of them independently. Thermodynamics deals with the system from the perspective of statistics of a collection of neutral particles and related the system energy directly to external stimulus. DFT considers an electron as a wavefunction with spatial distribution probability with involvement of limited degrees of freedom; MD takes a $H_2O$ molecule as the basic structural unit of polarizable or non-polarizable dipoles with attention more to the intermolecular O:H interaction, drift motion, spatial and temporal performance, and refers the O:H as the hydrogen bond that is incomplete. Integrating the inter- and intramolecular O:H–O cooperativity and nonbonding electron polarization would be much more appealing. Molecular motion dissipates energy caped at the O:H scale that is only less than 5% of the H–O energy. The H–O absorbs or emits energy through relaxation. Any detectible quantities are functional dependence on the physically elemental variables of length L, mass m, and time t (such as energy $[E] = [L^2/(mt^2)]$, frequency $[\omega^2] = [E/L^2]$, critical temperature for phase transition $[T_C] = [E]$, elasticity $[B] = [E/L^3]$, etc.) and their variations with external perturbation. Therefore, it would be more efficient to focus on the structural geometry and energy exchange of the O:H–O bond responding to perturbation, as the key driver of solvation study and molecular engineering and science.

8) Focusing on the bond-electron-phonon-property correlation and interlaying the spatially- and temporarily-resolved electron/phonon/photon spectrometrics would substantiate the advancement of related studies. Combining the spatially resolved electron/phonon DPS and the temporarily resolved ultrafast pump-probe spectroscopies not only distill the phonon abundance-stiffness-fluctuation due to the conditioning liquid but also fingerprint the electron/phonon energy dissipation and the ways of interactions. Molecular residing time or drift motion under a certain coordination environment fingerprints the way of energy dissipation but these processes could hardly give direct information of energy exchange under perturbation. Polarization, entrapment, or defect edge reflection and absorption determine the energy dissipation. Embracing the emerged O:H–O bond segmental disparity and cooperativity and the specific heat difference would be even more revealing.

Understanding may extend to water-protein interaction, biochemistry, environmental and pharmaceutical industries. Hydrophobic interface is the same to free surface. Charge injection by salt and other solute solvation provides the local electric fields. As the independent degrees of freedom, molecular undercoordination and electric polarization are ubiquitous to our daily life and living conditions. Knowledge developed could contribute to the science and society. It would be very promising for one to keep mind open and always on the way to developing experimental strategies and innovating theories toward resolution to the wonderful world.

## Acknowledgement

Financial support from the Natural Science Foundation (Nos. 21875024; 11872052) and National Science Challenge Project (No. TZ2016001) of China are all gratefully acknowledged.